# A Generic Library for Stencil Computations

Mauro Bianco[a], Ugo Varetto

*Swiss National Supercomputing Centre (ETHZ/CSCS)*

**Abstract**

In this era of diverse and heterogeneous computer architectures, the programmability issues, such as productivity and portable efficiency, are crucial to software development and algorithm design. One way to approach the problem is to step away from traditional sequential programming languages and move toward domain specific programming environments to balance between expressivity and efficiency.

In order to demonstrate this principle, we developed a domain specific C++ generic library for stencil computations, like PDE solvers. The library features high level constructs to specify computation and allows the development of parallel stencil computations with very limited effort. The high abstraction constructs (like **do_all** and **do_reduce**) make the program shorter and cleaner with increased contextual information for better performance exploitation. The results show good performance from Windows multicores, to HPC clusters and machines with accelerators, like GPUs.

Project ID: #283493

## 1. Introduction

The role of stencil computations is very important in many fields of scientific computing. Finite difference equations represent probably the most typical use of stencil computations, but there are others, like lattice methods, basic image processing filters, some dynamic programming algorithms, and others. In this paper we present GSCL (*Generic Stencil Computing Library*) [5] to let application programmers specify stencil computations at high level over regular grids using *generic programming* capabilities of C++.

Given the heterogeneity of current computer architectures, raising the level of abstraction can be a useful solution to exploit such diversity. Stating: "*for all elements of the grids apply the function operator*" is a very general way of expressing the need of the computation that can then be implemented specifically for each particular platform. Raising the level of abstraction is suitable for specific classes of computations to limit the complexity of such a language, in our case at a level of a C++ library. However, when proposing some high level concepts, the reactions are typically of skepticism, since *abstraction* and *control* are perceived as competing requirements.

To alleviate this phenomenon, we follow the approach of generic programming, in which there always exists a generic (possibly slow) implementation for any construct. If some other implementation is found that is better suited for the platform or the input, then it is picked up and used in the implementation. Ideally, the efficient implementation is developed by some library developer that may or may not be acquainted with the application domain of the final application. Furthermore by keeping GSCL architecture as simple as possible, we try to open the possibility for the user to implement, and possibly to contribute to the GSCL itself, with their own specific implementations. From the productivity point of view, this has the advantage that, if the proper specialization already exists the user can remain at the high level of abstraction and obtain the best sustained performance; while if this is not the case, the development is naturally split into two phases: First there is a quick deployment of the program, then increasingly more efficient versions can be developed. Those versions do not impact the high level code which can still be read and understood by the application programmers in a much better way than the low-level dirty-but-fast detailed implementation targeted to the architecture.

---

[a] Corresponding author: mbianco@cscs.ch

The diversity of the available platforms imposes certain restrictions in the high-level language that the GSCL supplies. We tried to keep the consequences of this to a minimum, but they have impact on the guaranteed semantics of the operations performed by a program. As an instance, limitations are dictated by the need of exploiting *accelerators* like GPUs, which are very attractive for stencil computations, since they have a distinct address space. These problems could be removed if GSCL was a stand-alone language. We choose to develop GSCL as a library, instead of a language, because of the possibility of using all the features of an existing language, like abstraction mechanisms, and the availability of a highly sophisticated, reliable, and widely adopted compiler technology.

**2. Related Work**

Stencil computations describe very important algorithms found in many applications. Lot of attention has been placed in studying stencil computation also because they are typically bandwidth limited, thus exercising the memory subsystem of computers.

In [2,8,13] the optimization of specific stencil computations are investigated which involves also time loops (i.e., the repetition of a stencil computation in an iterative fashion). Similar stencils are studied in [4,14] in the context of cache obliviousness. These efforts differ from GSCL since we focus on more complex stencil computations from the point of view of expressivity and incremental optimization. In the mentioned papers, the stencil computations implemented are at most a Jacobi iteration with two grids, while real applications like climate modeling or earthquake simulations [3,10] take 5 to 15 grids at once. It is not clear in what extent the reasoning behind many optimizations for simpler stencils could be directly ported to these cases.

In [6] auto-tuning techniques are studied to optimize performance of widely known stencil applications to compute gradient, Laplacian, and divergence. In [1] a language to express stencil computations is proposed, along with an auto-tuning framework to generate executable code highly optimized for a given architecture. Even though we agree that auto-tuning is very important, GSCL is implemented as a C++ library instead of a separate language with an associated code generation tool chain. It is our opinion that this would facilitate the adoption of GSCL and also the integration with existing programs. We plan to adopt auto-tuning techniques in GSCL, but this is out of the scope of this report. Similarly [7] implements a high level interface in C, which however makes extensive use of macros, which can lead to maintainability issues. In GSCL we try to keep the use of macros to the minimum.

In the context of grid computing [11,12] explore the possibility of raising the level of abstraction using a C++ generic library to express the computation in presence of high latency networks. The authors exploit the mechanism of using wide ghost areas to hide latency, which is under investigation in GSCL in the contexts of GPUs.

**3. Programming in GSCL**

In GSCL we define a stencil computation as follows. We define a *regular grid* (in the following it will be called simply *grid*) G as a *d*-dimensional array accessed using the notation $G(i_0, i_1, \ldots i_{d-1})$.

A *stencil operator* is a function (intended in C/C++ sense, so it's not a function in mathematical sense) that takes a tuple $(i_0, i_1, \ldots, i_{d-1})$, called the *core element*, and a tuple of grids $(G_0, G_1, \ldots, G_k)$, and *reads* elements of the grids at *fixed offsets* from the core element in each grid and *can write* elements corresponding to the core elements of the grids. A stencil operator can also return a value as result of the computation.

A *stencil computation* is the application of a stencil operator to a partially ordered set of tuples $\{(i_0, i_1, \ldots, i_{d-1})_j: j=0\ldots M-1\}$, which, in GSCL, are called *Iteration Spaces*. Iteration spaces require the list of grids to process as inputs and the stencil operators to be applied. The iteration spaces available in current state of GSCL are:

- **do_all**: does not guarantee any order of application of the stencil operator;

- **do_reduce**: does not guarantee any order of application of the stencil operator but the return value of the stencil operators are (commutatively) reduced to a single value through a user defined reduction operator;
- **do_diamond**: cells (i-1,j) and (i,j-1) are processed before cell (i,j) (available only for 2D at the moment);
- **do_i_inc**, **do_j_inc**, **do_k_inc**: cells (i-1, j, k), (i,j-1,j), (i,j,k-1), respectively, are processed before cell (i,j,k);
- **do_i_dec**, **do_j_dec**, **do_k_dec**: cells (i+1, j, k), (i,j+1,j), (i,j,k+1), respectively, are processed before cell (i,j,k);

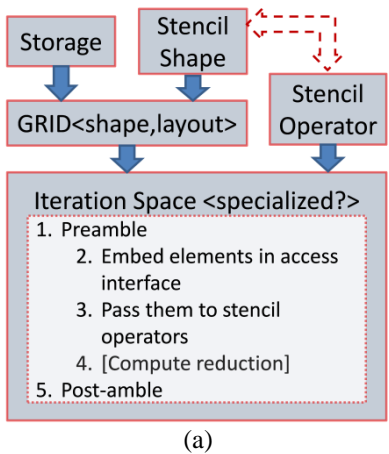

```
struct stencil_operator {
  typedef access_list<flag_write,flag_read>
                                     access_list_type;

  template <class S1, class S2>
  void operator()(S1 &v, S2 const& u) const {
  v() = 1.0/36.0 *(6*u()
                          - u(1,0,0) - u(-1,0,0)
                          - u(0,1,0) - u(0,-1,0)
                          - u(0,0,1) - u(0,0,-1));
  }
};
```

(a)                                      (b)

**Figure 1** On the left, the sketch of the software architecture diagram of the library. On the right, an example of stencil operator with some traits to customize it.

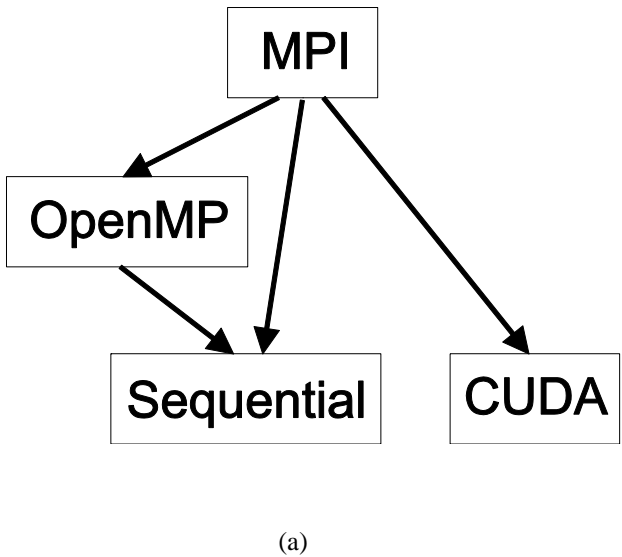

```
// behavior here is undefined
GSCL_Init(argc, argv);
// Here init of storage and grids
context<arch> ctx = GSCL_Begin<arch>();
// Next statement is OK: Context is active
do_all(ctx, grid, stencil_operator);
// Next is wrong
cout << grid(10,10) << "\n";
GSCL_End(ctx);
// Next is OK value is printed
cout << grid(10,10) << "\n";
// Next is ERROR: context invalid
do_all(ctx, grid, stencil_operator);
GSCL_Finalize();
// Behavior here is undefined
```

(a)                                      (b)

**Figure 2** On the left (a) the architecture layers used to describe the required architecture. On the right (b) a sample code to show semantics of operations.

Figure 1.a shows the main components of GSCL and their interactions are depicted. In GSCL, grids are made by the conjunction of a storage object and a stencil shape. The stencil shape serves two purposes. First it defines the *halo* the stencil operators are allowed to access around the core element, second, they provide the interface to access elements in stencil operators. In Figure 1.b, a stencil operator is shown which uses interfaces like **v()**, **u(0,-1,0)**, and so on, to access elements at core element or around it that are defined in the stencil shape class. GSCL provides two types of stencils shapes. One type is called *stateful* stencils, and these allow the user to query the indices of the accessed elements. This is useful, for instance, for initialization and implementing red-black Gauss-Siedel algorithm. The other type of stencil shape cannot perform those operations. Users can define their own shapes.

As shown in Figure 1.b, stencil operators are C++ function objects (functors) with some specific characteristics. The most important is that its function operator must be a template. Additional optional *traits* can be added to the stencil operators to convey additional information to the GSCL, as, for instance the access mode (read-only, write-only, etc.) type for each grid in input, which is very important for performance reasons.

Given a GSCL program, the implementation of the stencil computations is chosen by the user through the use of architecture types. GSCL uses a hierarchical description of the platform/architecture in hand. Each level of the hierarchy is the *programming model/paradigm* at that level. The hierarchy is described as a DAG, as in Figure 2.a. The programmer can pick one model in each level, starting anywhere in the DAG and construct an *architecture*. As an example, suppose the machine is made of nodes that communicate through MPI, each node uses multi-threads through OpenMP, and each thread executes a regular sequential code. In such a machine we can define the hierarchy by using **GSCL::make_arch**:

**typedef make_arch<mpi,openmp,sequential>::type a_type;**

Once **a_type** is defined, this is used to instantiate execution contexts, which are passed around in GSCL, thus informing about the programming model active at a given stage. Not all the combinations are supported at the moment, for instance **GSCL::cuda**, for running on GPUs, can only be used as top model or under **GSCL::mpi**.

Changing architecture will be a matter of changing the lines of code that define **a_type**s. Although this requires a change in very few lines of code (typically one), the program is no longer strictly portable. To fix this problem, GSCL provides the **default_arch** type, which gives a predefined hierarchy to exploit the architectural features available on the machine. Depending on the compiler used, the default architecture can be printed as *pragma* information during compilation in order to check the correctness of the environment.

Data storage classes match with the architecture hierarchy. The use of defaults turns out to be useful here. The library uses a parameterized default storage that select the storage implementation based on the architecture type

**typedef default_Storage<arch_type, T>::type sbstorage_t;**

Where **T** is the type of the values stored in the storage object. Being parameterized with architecture type, the storage type does not need to change when the code is ported to another machine.

## 4. Designing for Portable Performance

One major source of issues for high level abstract constructs is *accelerators*. This category includes GPUs, nominally capable of high throughput that is the main characteristic needed by a stencil computation. Since GPUs and CPUs access different address spaces, users need to have some semantic information about when certain data can be accessed and where. For this reasons we introduced the concept of *context*. A context is obtained through a call to **GSCL_Begin** and has a template argument that is the architecture type described in the previous section. The semantic of the operation, as seen in Figure 2.b, is that, in between **GSCL_Begin** and **GSCL_End** statements, direct access to the data, through grid and storage interfaces, lead to unpredicted results. Grids can only be passed as arguments to iteration spaces. After an **end** statement and before the next **begin**, data can be accessed, but GSCL calls to iteration spaces abort the program with an error message. This precaution makes it possible to keep data on the GPUs memory between invocations of GSCL constructs, thus amortizing the high latency of these devices.

GPU support in GSCL is implemented in CUDA [9] since it is the only available environment that supports a subset of C++ for on-device computation.

## 5. Evaluation

The most important test for GSCL at this stage is the evaluation of the *cost of abstraction*, that is, the loss of performance, if any, due to the fact that we are giving the programmer a high abstraction interface that does not provide (at first) control on implementation. Next subsection describes the machines used during evaluation.

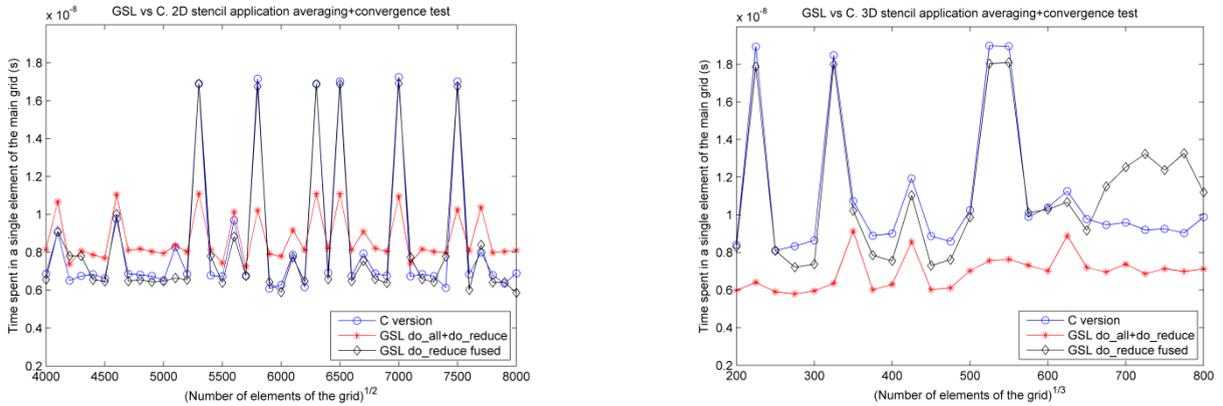

**Figure 3** Comparison of GSCL and C99 code. The fused version is algorithmically equivalent to the C version, while the other does an additional scan of the memory since the loops are not fused.

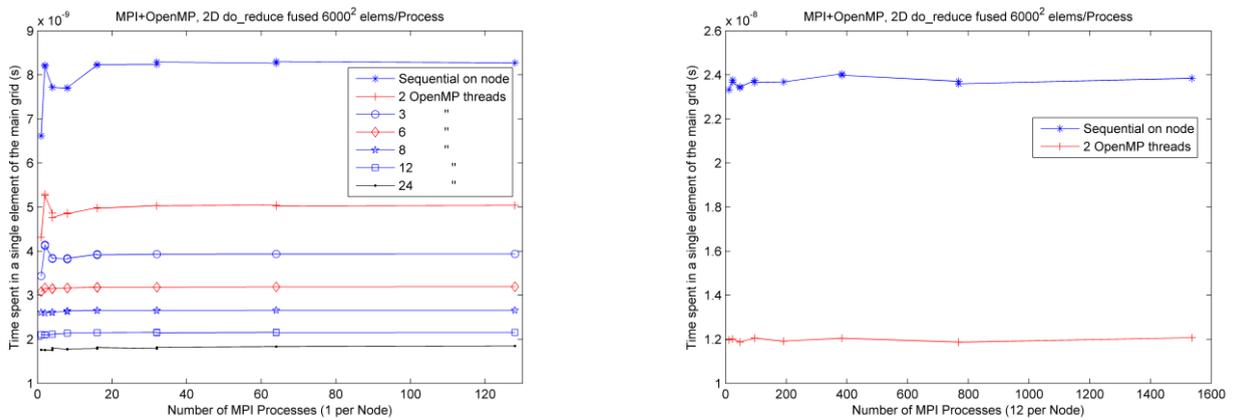

**Figure 4** Running GSCL code on a Cray XT5 machine. Each node has 24 cores divided in two sockets. On left one MPI process per node is used and then a varying number of OpenMP threads for process are used. On the right the node is half filled with MPI processes and we are showing none or 2 OpenMP threads

### 5.1. Description of the machined

The comparison of GSCL against C and the MPI and OpenMP versions are executed on CRAY XT5 with 176 nodes available at CSCS. Each node has two 12-core AMD Opteron CPUs running at 2.1GHz, for a total of 24 cores. The interconnect between nodes is a 10.4 GB/s 2 Gemini.

The results for GPUs have been executed on a 12 cores node (two 6 cores chips) at 2.2GHz. The board carries two nVidia M2090 GPU cards. The GPU cards run at 1.2GHz and have 6GB of memory each.

### 5.2. Cost of abstraction on CPUs

To test this we ran a very simple application that iteratively applies an averaging stencil on a grid, writing the results in another grid (Jacobi iteration) and stops when all elements computed in a new iteration are close to the ones of the previous iteration (infinite norm). The application does a **do_all** to apply the averaging on the main grid, and then executes a **do_reduce** to check the current values with the previous ones. In this example we use the same operator showed before in Figure 1.b for averaging.

This may not be the best solution, especially when there are few grids, like in this case. We can fuse the two operators together and obtain the following code for the main loop:

```
do
  swap_grids(); // to update which has to be read
  res = do_reduce(context, grid_now, grid_before,
                  fuse(sten_op_diffusion(),sten_op_convergence(EPSI)),
                  std::logical_and<bool>());
while (!res)
```

In this case only one scan of the grids is needed. An important aspect is that a same GSCL code will run sequentially, on OpenMP shared memory nodes, MPI clusters with and without OpenMP nodes, on GPUs, on MPI with GPUs. The design of GSCL stresses the easy extension to future other platforms, without changing a single line of application code, if the defaults described previously are used. The C programmer must adapt the code to the architecture, thus affecting the code readability and portability. The C code, algorithmically equivalent to the fused version, runs only sequentially in this example.

The first test we show is then comparison of GSCL and C on the above described application for different input sizes. Figures 3.a and 3.b shows the results for 2D and 3D cases, respectively. The given times are the average time spent on an element of the grid. It is computed by taking the overall time and dividing it by the number of elements of the main grid and by the number of iterations. This is useful to compare different input sizes.

As can be seen in Figure 3.a, in the 2D case, the GSCL version with a single fused reduction is on par with the C version. The peaks at certain given sizes are due to memory architecture issues. It is interesting to note that the implementation with two distinct iteration spaces has a more stable behavior than the other two, even though it tends to be slower in the best case. The reason is because the cache of the system is less stressed by splitting the operations, thus behaving more smoothly.

In Figure 3.b the results shows that GSCL is still the fastest, but the two iteration spaces implementation is actually faster than the others. This is again due to memory pressure since the stencil operator accesses 6 elements at different strides instead of 4 in the 2D case. The two algorithmic equivalent versions are still on par up to a 6400x6400 input size. After that the C++ version exhibits slightly higher execution times. We would like to stress that, in any case, all the optimizations available for the C version could be applied to the GSCL versions, with the advantage that main application code is not affected. Also, the difference between the two iteration spaces and the fused operators versions differ by only a couple of lines of code.

To prove that GSCL can run efficiently on traditional parallel machines, we took the code that ran the comparison with the C implementation, and compiled using MPI and OpenMP. We performed a weak scaling experiment in which each MPI process had a tile of 6000x6000 elements. The shape of the problem grid is the one with the aspect ratio closest to 1 given the number of processes. Results shown in Figure 4.a place a single MPI process in a node of the machine. The process then employs from 1 to 24 openMP threads to work on the tile. As can be seen, the time with one OpenMP thread is the same as the sequential time shown in Figure 3.a. Increasing the number of OpenMP threads the execution time improves according to our experience on the machine.

Figure 4.b shows the same code executed by placing 12 MPI processes on each node. As we expect, the execution time increases due to contention of the many processes. The weak scalability of the application is, however, very good.

*5.3. Cost of abstraction on CPUs*

**Table 1** Times of the kernels in GSCL and CUDA. Data transfer time not included to test the cost of abstraction due to GSCL.

| Operation | GSCL | CUDA |
|---|---|---|
| **do_all** | 3m | 4ms |
| **do_reduce** | 10ms | 9ms |

The cost of abstraction on GPUs are evaluated slightly differently. First we measured the time of a simple **do_all** and a **do_reduce** without taking into account communication between CPU and GPU. The stencil operator performs the usual averaging, while the reduction is computing the sum of all elements. The results are shown in Table 1. As can be seen GSCL is faster than the corresponding CUDA C version for **do_all** and 10\% slower for the reduction.

# 6. Conclusion

In this report we presented GSCL, a generic programming C++ library that raises the level of abstraction of stencil computations. This allows application programmers to concentrate first on problem specific aspects instead of low level details that are relevant for performance. The structure of the library allows for incremental improvement of performance either by the application programmer or the library developer. We showed that the cost of abstraction is very low and that GSCL achieves the same level of performance of corresponding C implementations on the same algorithms in sequential, MPI and OpenMP parallel implementations, and for GPUs.

In the future we plan to extend GSCL to even more architectures and programming models, and produce more specializations for existing cases. We also plan to introduce autotuning in GSCL. Also, we plan to demonstrate the effectiveness of GSCL in state of the art scientific codes, such as seismic simulations and lattice methods for fluid dynamics.

# Acknowledgements

This work was financially supported by the PRACE project funded in part by the EUs 7th Framework Programme (FP7/2007-2013) under grant agreement no. RI-211528 and FP7-261557. The work is achieved using the PRACE Research Infrastructure resources at Swiss National Supercomputing Centre (CSCS) in Switzerland.

# References

[1]   M. Christen, O. Schenk, and H. Burkhart. Automatic code generation and tuning for stencil kernels on modern shared memory architectures. Comput. Sci., 26:205– 210, June 2011.

[2]   K. Datta, S. Kamil, S. Williams, L. Oliker, J. Shalf, and K. Yelick. Optimization and performance modeling of stencil computations on modern microprocessors. SIAM Rev., 51:129–159, February 2009.

[3]   G. Doms and U. Schatter. A description of the nonhydrostatic regional model lm, part i, dynamics and numerics. http://cosmomodel.org/content/model/documentation/core/cosmoDyncsNumcs.pdf, 2002.

[4]   M. Frigo and V. Strumpen. Cache oblivious stencil computations. In Proc. of the $19^{th}$ Int. conf. on Supercomputing, ICS '05, pages 361–366, New York, NY, USA, 2005. ACM.

[5]   GSCL Web Site: https://hpcforge.org/projects/stencilplusplus/

[6]   S. Kamil, C.Chan, L. Oliker, J.Shalf, and S. Williams. An auto-tuning framework for parallel multicore stencil computations. In IPDPS, IPPS 2010, pages 1–12, 2010.

[7]   N. Maruyama, T. Nomura, K. Sato, and S. Matsuoka. Physis: an implicitly parallel programming model for stencil computations on large-scale gpu-accelerated supercomputers. In Proc. of 2011 Int. Conf. for High Performance Computing, Networking, Storage and Analysis, SC '11, pages 11:1–11:12, New York, NY, USA, 2011. ACM.

[8]   A. Nguyen, N. S*atish, J. Chhugani, C. Kim, and P. Dubey. 3.5-d blocking optimization for stencil computations on modern cpus and gpus. In Proc. of the 2010 ACM/IEEE Int. Conf. for High Performance Computing, Networking, Storage and Analysis, SC '10, pages 1–13, Washington, DC, USA, 2010. IEEE Computer Society.

[9]   Nvidia. Nvidia cuda compute unified device architecture, 4.0 edition, 2011.

[10]   O. Rojas, E. M. Dunham, S. M. Day, L. A. Dalguer, and J. E. Castillo. Finite difference modelling of rupture propagation with strong velocity-weakening friction. Geophysical Journal International, 179(3):1831–1858, 2009.

[11]   A. Schafer and D. Fey. Libgeodecomp: A grid-enabled library for geometric decomposition codes. In Proceedings of the 15th European PVM/MPI, pages 285–294, Berlin, Heidelberg, 2008. Springer-Verlag.

[12]   A. Schafer and D. Fey. High performance stencil code algorithms for gpgpus. Procedia CS, 4:2027–2036, 2011.

[13] T. Shimokawabe, T. Aoki, T. Takaki, T. Endo, A. Yamanaka, N. Maruyama, A. Nukada, and S. Matsuoka. Peta-scale phase-field simulation for dendritic solidification on the tsubame 2.0 supercomputer. In Proceedings of 2011 International Conference for High Performance Computing, Networking, Storage and Analysis, SC '11, pages 3:1–3:11, New York, NY, USA, 2011. ACM.

[14]   R. Strzodka, M. Shaheen, D. Pajak, and H.-P. Seidel. Cache oblivious parallelograms in iterative stencil computations. In Proceedings of the 24th ACM International Conference on Supercomputing, ICS '10, pages 49–59, New York, NY, USA, 2010. ACM.